\documentclass[aps,rmp,reprint]{revtex4-1}
\usepackage{graphicx}
\usepackage{color}      
\usepackage{amsmath}
\usepackage{textcomp}
\usepackage{hyperref}

\newcommand{\vect}[1]{\mathbf{#1}}

\newcommand{\citeasnoun}[1]{\citet{#1}}
\newcommand{\ack}[1]{\begin{acknowledgements}#1\end{acknowledgements}}
\newcommand{\referencelist}[1][iucr]{
\bibliography{#1}%
}
\begin{document}
\title{A diffraction effect in X-ray area detectors}

\author{Christian Gollwitzer}
\email[Correspondence e-mail: ]{christian.gollwitzer@ptb.de}
\author{Michael Krumrey}
\affiliation{Physikalisch-Technische Bundesanstalt (PTB), Abbestr. 2-12, D-10587 Berlin, Germany}

\begin{abstract}
When an X-ray area detector based on a single crystalline material, for instance, a state of the art hybrid pixel detector,
is illuminated from a point source by monochromatic radiation, a pattern of lines appears which overlays the detected image.
These lines can be easily found by scattering experiments 
with smooth patterns, such as small-angle X-ray scattering. The origin of this effect is the Bragg reflection 
in the sensor layer of the detector.
Experimental images are presented over a photon energy range from 3.4\,keV to 10\,keV, together with a
theoretical analysis. The intensity of this pattern is up to $20\%$, which can
disturb the evaluation of scattering and diffraction experiments.
The patterns can be exploited to check the alignment of the
detector surface with the direct beam, and the alignment of individual detector modules with each other in the case of 
modular detectors, as well as for the energy calibration of the radiation.
\end{abstract}

\maketitle

\section{Introduction} 
\label{sec:introduction}

Images of X-ray scattering and diffraction experiments are typically recorded nowadays by using a large-area digital
detector. The recent advances of so-called hybrid pixel detectors, where an electronic circuit board
is bump bonded to a sensor layer \cite{heijne1989}, 
have improved the image quality to a large extent. State-of-the art
detectors like
the XPAD~\cite{delpierre2007},
detectors based on the Medipix read-out chip~\cite{ponchut2001,pennicard2010}, and the 
PILATUS \cite{kraft2009,donath2013} combine a pixelated semiconductor sensor with a circuit
board providing an amplifier, a discriminator and a digital counter for every single pixel. By means of this
technique, the detector directly counts single X-ray photons for every pixel, and the image quality
in terms of the signal-to-noise ratio
is ultimately only limited by the quantum noise of the photons. In addition, large dynamic ranges
and very small crosstalk between neighbouring pixels can be achieved. 

The sensitive layer of such a detector, which absorbs the photons and converts them into an electric signal,
is usually made of a single crystal silicon wafer, but CdTe or CdZnTe
sensors are also in use for detectors operating in the $10\,$keV to $500\,$keV photon energy range \cite{takahashi2001}. 

Mostly, the
crystalline structure of the sensor layer of the detector is of minor importance to the primary
scattering experiment. However, Bragg reflection in this layer, which is characteristic of a crystalline material,
can significantly change the detected signal. This was shown by \citeasnoun{zheludeva1985}
for a photodiode point detector and by \citeasnoun{holy1985}, who measured the photoconductivity of the
second crystal in a double-crystal monochromator as a function of the angle.
The strong angular and energy dependency of the detected signal as a result of the Bragg
diffraction was first exploited by \citeasnoun{jach1988}, who embedded a photodiode directly in the 
monochromator crystal to assist in tuning the crystals \cite{jach1990}.
Later, \citeasnoun{erko2001} and \citeasnoun{krumrey2001} used the effect in an external photodiode
to calibrate the energy scale of their X-ray monochromators. \citeasnoun{hoennicke2004} implemented
it as a novel method to detect a diffracted beam at an angle of 90\textdegree. 

\citeasnoun{hoennicke2005} extended the idea of \citeasnoun{jach1988} by using a CCD as a 
monochromator crystal to simultaneously monochromatize the incoming X-rays and 
detect a spatially resolved image. They observed a dark curved line 
across the detected image which depends on the photon energy and the angle of incidence and
indicates the position on the crystal at which the Bragg condition is fulfilled.

\begin{figure}
\includegraphics[width=0.8\columnwidth]{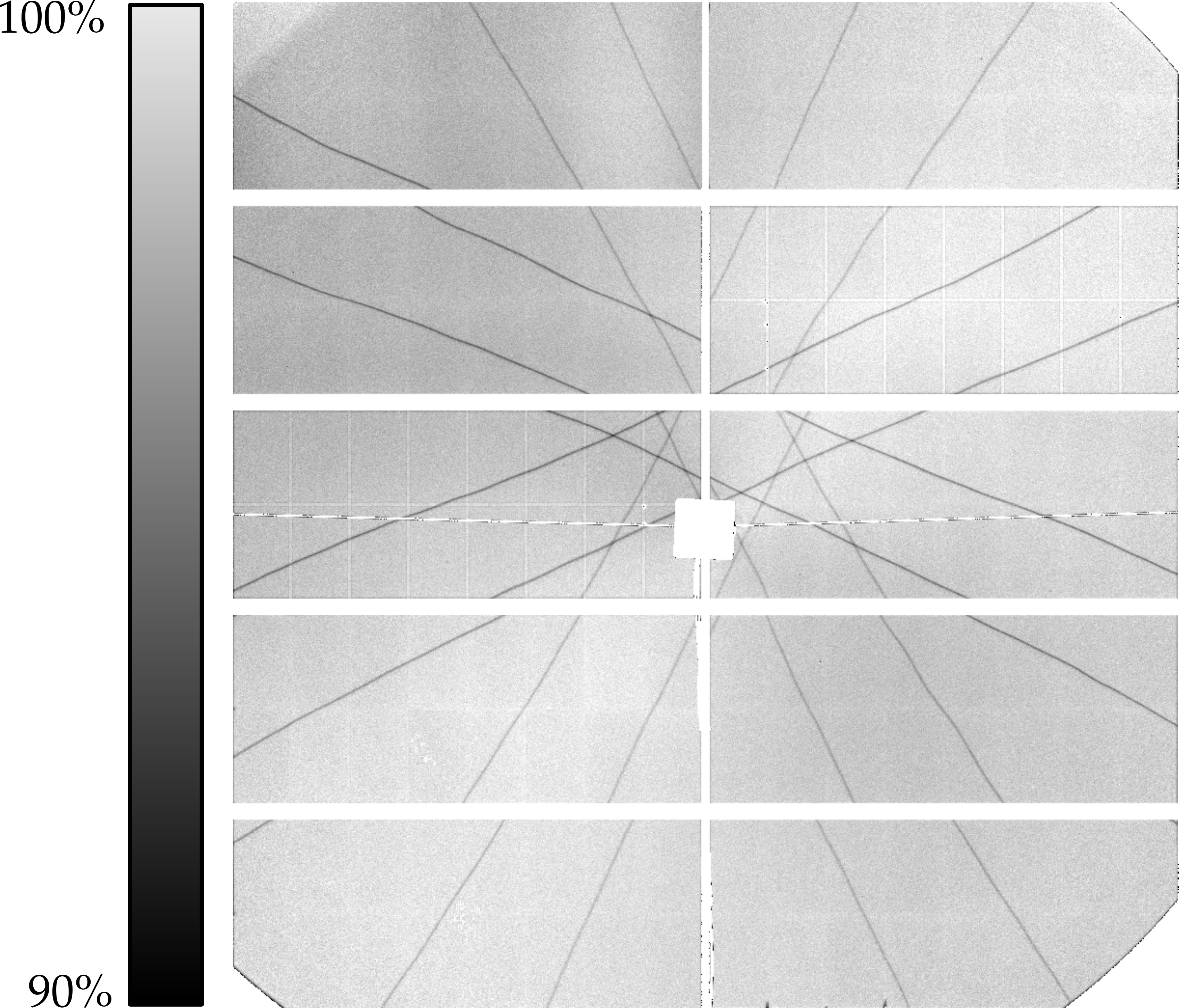}
\caption{Example of a pattern observed with a hybrid pixel detector, nearly homogeneously
illuminated by a monochromatic point source  with a photon energy of $E_\text{ph}=7970$\,eV 
at a distance of $d=2333\,\mathrm{mm}$. The incident intensity varies from $10^5$ photons per pixel in the
centre to $3\times10^4$ photons in the edges of the image, collected over a total recording time of $52$
minutes. The grey scale denotes the ratio of registered to expected counts in every pixel. The detector covers an area of
$170\times180\,\mathrm{mm^2}$. Masked regions, e.g.\ the beamstop, gaps 
between detector modules, and shadowed areas, are displayed in white.}
\label{fig:example}
\end{figure}

A similar pattern can arise in every diffraction experiment, when a single crystalline imaging
detector is used. In this paper, regular patterns in the detected images of a photon-counting
hybrid pixel detector are reported which are caused by elastic Bragg scattering in the sensor layer of
the detector. Figure~\ref{fig:example} displays an example image recorded by a homogeneously
illuminated detector. The image was high-pass filtered to enhance the visibility of the pattern 
using the procedure described in detail later in this paper. 
Ideally, this image would be completely featureless, but the Bragg pattern 
overlays the detected image in the form of faint dark lines. 
When the energy is changed, the lines move across the entire image and finally disappear when
they move out of the detector area.

These artefacts are always present when the detector
is illuminated by monochromatic radiation from a point source, which is the
typical configuration for many scattering and diffraction experiments. However, for primary signals
with a large variation in contrast and small-scale features such as crystallography diffraction images, the
patterns may go unnoticed. They are most easily
observed for small-angle scattering (SAXS) experiments which can lead to smooth scattering patterns. 

The formation of these patterns can be explained as
follows: at every detector pixel, the angle of incidence of the incoming photons is different in the
crystalline coordinate system. At some point, the incident photon may fulfill the Bragg condition of
an arbitrary symmetry plane. This photon has a finite probability to be elastically reflected out of
the sensor layer, and consequently is not measured. This leads to a lower signal of the
detector at the corresponding pixel, and possibly a higher signal at a nearby pixel, if the photon is
still absorbed inside the detector. The mechanics of losing a
photon due to parasitic Bragg scattering resembles the well-known monochromator
glitches \cite{vanderlaan1988}. When the photon energy is fixed, the Bragg condition still
leaves one degree of freedom for the incident direction, namely the rotational symmetry about the
normal vector of the Bragg plane. It follows that the pattern consists of lines.

This paper is organized as follows. First, the experiments are described in detail.
Then a theoretical analysis of the patterns is given based on Bragg diffraction. 
Next, the experimentally observed images are compared with the theoretical predictions 
and finally possible applications of the phenomenon are discussed.

\section{Experiments and data processing}
\begin{figure}
\includegraphics[width=0.9\columnwidth]{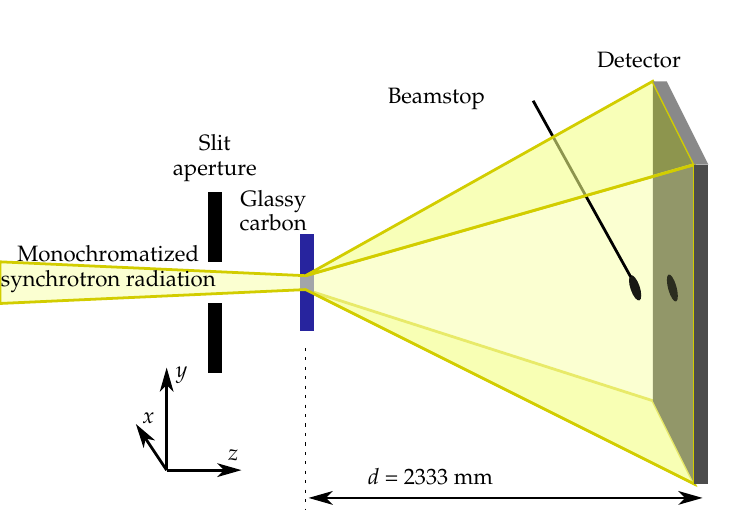}
\caption{Experimental setup to record the Bragg patterns}
\label{fig:setup}
\end{figure}

The experimental setup is illustrated in figure~\ref{fig:setup}.
It is based on a common small-angle X-ray scattering setup at the synchrotron radiation
facility BESSY~II at the FCM beamline~\cite{krumrey1998,krumrey2001}.
The radiation from a bending magnet is monochromatized using
a four-crystal monochromator equipped with InSb(111) and Si(111) crystals, 
which provides an energy range from $1.75$\,keV to $10$\,keV. 
The monochromatized radiation is then collimated using a slit aperture of the size 
$1.2\times1.2\,\mathrm{mm^2}$ and
focused on the sample. The key point in observing the patterns is the
selection of a sample which scatters uniformly over the whole detector area. For the energy range
from $6\,\mathrm{keV}$ to $10\,\mathrm{keV}$, a $1\,\mathrm{mm}$ thick sample
of glassy carbon was used, and below $6\,\mathrm{keV}$, a sample of the same material with a thickness of
$90\,$\textmu m was used. Glassy carbon is a common reference material in SAXS
due to its low variation in scattering intensity over the range of the momentum transfer $q$ between
$0.1\,\mathrm{nm^{-1}}$ and $1\,\mathrm{nm^{-1}}$~\cite{zhang2010}, 
making it an ideal scatterer to provide a nearly homogeneous illumination. 

The scattered radiation is then collected by a PILATUS~1M detector from Dectris Ltd., modified
for direct operation in vacuum~\cite{donath2013,wernecke2013pil}. This detector consists of ten modules, each of
which contains $487\times195$ photon counting pixels with a pixel size of $172\,$\textmu m.
The sensor layer of each module is made of a silicon
wafer with the $(001)$ plane facing the beam. The long side of the module is parallel to the
$[110]$ axis of the wafer, which corresponds to the detector $x$-axis.
This detector is mounted on the SAXS setup of the Helmholtz-Zentrum Berlin~\cite{krumrey2011}, 
which allows positioning with \textmu m resolution along the beam ($z$ axis in figure~\ref{fig:setup})
and perpendicular to it ($x$ and $y$ axes).
The sample-detector distance was set to $d=2333\,\mathrm{mm}$ for all experiments.

The images were subsequently filtered using the following
procedure to enhance the visibility of the patterns. First, an azimuthal averaging is performed, which is a
standard preprocessing step for SAXS data \cite{pauw2013}.
During this processing, all pixels which deviate by more than 3 standard deviations from the median intensity are treated as
outliers and are excluded.
Then, the original data is divided by the resulting scattering
curve pixel by pixel. This results in an image of the relative deviation of every pixel from the
mean value for the corresponding $q$. This processing acts as a
(non-linear) high-pass filter with the corresponding low pass defined by the azimuthal averaging step.
Finally, the contrast range of the image is adjusted for visualization.
A range of $90\,\%$ to $102\,\%$ of the full scale intensity is suitable for most of the patterns.

The image in figure~\ref{fig:example} was processed in this way to clearly display the pattern.
The raw image varies in intensity by more than a factor of 3 from the centre to the border, because the
scattering from glassy carbon is not perfectly isotropic. The pattern, on the other hand,
changes the intensity by only $3.5\%$ and is hard to detect by eye when the full dynamic range of the image is
displayed. The filter makes the pattern easily visible across the whole image
and the full energy range by removing the background. 

Some regions of the original image have been masked after the processing. 
This includes the beamstop, together with the mounting parts, the gaps between the individual 
detector modules, and areas shadowed by elements of the beamline. 
These areas are marked in the filtered images in white.

\section{Theory} 

\label{sec:theory}
\begin{figure}
\includegraphics[width=0.8\columnwidth]{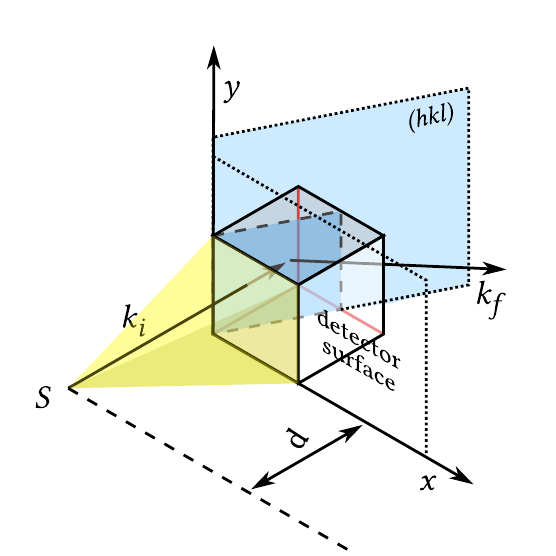}
\caption{Geometry of the scattering problem. The cube represents one pixel of the detector.
The photons emanate from the monochromatic point source
at $S$, hit the detector surface at $(x,y,d)$ and are reflected at the plane $(hkl)$ displayed in
light blue.}
\label{fig:geometry}
\end{figure}

The geometry of the problem is depicted in figure~\ref{fig:geometry}. The detector is illuminated
from a monochromatic point source located at $S$ at a distance $d$ from the surface. 
Consider a crystallographic plane with the Miller indices $(hkl)$, which is generally
not parallel to the detector surface. A pattern line can appear at every position at the surface, where the incident
photon with the wave vector $\vect{k_i}$ fulfills the Bragg condition
\begin{equation}
2\vect{k_i}\cdot\vect{G}=G^2,
\label{eq:Laue}
\end{equation}
where $\vect{G}=[hkl]$ is the reciprocal lattice vector of the corresponding plane.
With the coordinates $(x, y)$ of the intersection of the detector surface and the ray,
the wave vector can be expressed in the experimental frame as
\begin{equation}
\vect{k_i}=\frac{2\pi}{\lambda}\cdot\frac{\left(
\begin{array}{c}
x \\ y \\ d
\end{array}
\right)}{\sqrt{x^2+y^2+d^2}},
\label{eq:wavevect}
\end{equation}
where $\lambda$ is the wavelength of the photon.
In order to substitute \eqref{eq:wavevect} into \eqref{eq:Laue}, the wave vector must be transformed
into the crystalline coordinate system. In this case the detector surface was
parallel to the crystallographic plane $(001)$, and the detector $x$-axis was 
parallel to the $[110]$ unit vector of the cubic lattice, which defines the coordinates of 
any point on the surface as
\begin{align}
x'&=\left(x+y\right)/\sqrt{2}\\\nonumber
y'&=\left(x-y\right)/\sqrt{2}\\\nonumber
z'&=d.
\end{align}

Substituting the transformed coordinates into the Laue equation \eqref{eq:Laue} yields the implicit function 
\begin{equation}
F(x',y')=\frac{hx'+ky'+ld}{\sqrt{x'^2+y'^2+d^2}} - \frac{\lambda}{2a}\left(h^2+k^2+l^2\right)=0,
\label{eq:pattern}
\end{equation}
where $a$ is the lattice constant of the cubic lattice.

The solution to this equation in the unknowns $(x', y')$ 
describes the expected patterns resulting from diffraction at the plane $(hkl)$.
Since the Bragg angle is invariant under the exchange of $h$ and $k$ and sign change, one has to respect
all these permutations when solving the condition~\eqref{eq:pattern}. This explains the 8-fold symmetry
of the patterns observed (cf.\ figure~\ref{fig:example}). If either $h$ or $k$ is zero, or $h=k$,
the symmetry collapses into a 4-fold pattern. If both $h$ and $k$ are zero,
equation~\eqref{eq:pattern} describes a circle with the radius
\begin{equation}
r=d\sqrt{\left(\frac{2a}{\lambda l}\right)^2-1}.
\end{equation}
Generally, equation~\eqref{eq:pattern} can be rearranged into a quadratic polynomial over
the detector coordinates $(x', y')$. Thus, all the lines represent conic sections.

For a given set of indices $(hkl)$, a characteristic wavelength or photon energy $E_{hkl}$ can be
defined where all lines of a specific pattern meet in the detector centre, at which point the incident photon 
angle is normal to the detector surface. The condition for this characteristic energy is given by 
substituting $x=y=0$ into equation~\eqref{eq:pattern} and solving for the energy
\begin{equation}
E_{hkl} = \frac{hc}{2a}\cdot\frac{h^2+k^2+l^2}{l}.
\label{eq:ecrit}
\end{equation}
In the case of the circular pattern, the condition~\eqref{eq:pattern} can only be fulfilled for photon
energies above $E_{hkl}$, at which the pattern consists of a singular point in the detector centre. 

In the next section, the numerical solutions to \eqref{eq:pattern} will be compared to the
experimental findings.

\section{Results and Discussion} 
\label{sec:results}
\begin{table}
\caption{Pattern indices sorted by characteristic energies for silicon $(001)$, observed
intensities and symmetry in the experimentally covered range from $3.4\,$keV to $10\,$keV. The sign
of the line intensity denotes dark, light, and dark/light doublet lines by $+$, $-$, and $\pm$,
respectively. The $\bigcirc$, $+$, $\times$, and $\ast$ signs denote circular, 4-fold upright, 4-fold diagonal 
and 8-fold symmetry, respectively.
For the silicon lattice constant, $a=543.108\,\mathrm{pm}$ was used.}
\label{tbl:ecrit}
\begin{ruledtabular}
\begin{tabular}{llllrc}
$h$ & $k$ & $l$ & $E_{hkl}$ (eV) & Intensity & Symmetry\\
1       &1      &1      &3424.29	&$+1.5\,\%$		&$+$\\
1       &1      &3      &4185.25	&$-5\,\%$		&$+$\\
2       &0      &2      &4565.73	&$-3\,\%$		&$\times$\\
0       &0      &4      &4565.73	&$-20\,\%$		&$\bigcirc$\\
1       &1      &5      &6163.73	&$-10\,\%$		&$+$\\
2       &2      &4      &6848.59	&$-7\,\%$		&$+$\\
3       &1      &3      &7229.06	&$\pm1\,\%$		&$\ast$\\
2       &0      &6      &7609.54	&$-8\,\%$		&$\times$\\
3       &1      &5      &7990.02	&$-3.5\,\%$		&$\ast$\\
1       &1      &7      &8316.14	&$-4\,\%$		&$+$\\
4       &0      &4      &9131.45	&$\pm0.5\,\%$	&$\times$\\
0       &0      &8      &9131.45	&$-5\,\%$		&$\bigcirc$\\
3       &1      &7      &9620.63	&$-2\,\%$		&$\ast$\\
3       &3      &5      &9816.31	&$-1.5\,\%$		&$+$\\
\end{tabular}
\end{ruledtabular}
\end{table}

Three different sets of scattering images from glassy carbon were recorded over the range from
$3.4\,$keV to $10$\,keV. The first set comprises an image at each of the characteristic energies
listed in table~\ref{tbl:ecrit}. The second set of images was recorded in steps of $5\,$eV 
over a photon energy range from $7.5$\,keV to $9.2$\,keV. The last set of images comprises a sequence
in steps of $\Delta E=0.1\,\mathrm{eV}$ and $\Delta E=0.2\,\mathrm{eV}$
at about the characteristic energies of the circular patterns $E_{004}$ and $E_{008}$, respectively.

\begin{figure}
\includegraphics[width=0.8\columnwidth]{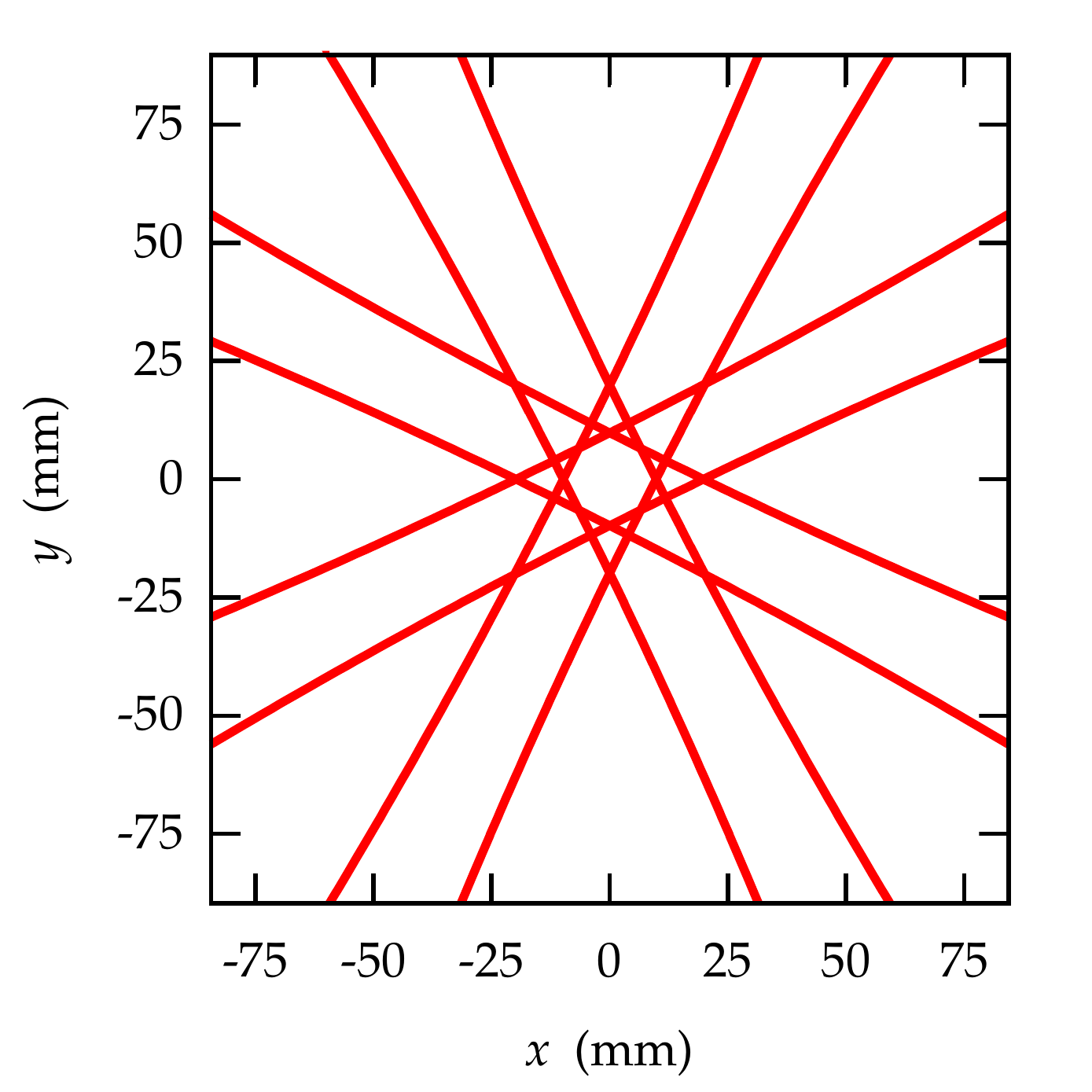}
\caption{Numerical solution of the Bragg condition~\eqref{eq:pattern} for a silicon crystal with the $(001)$ 
plane facing the beam. The Miller indices were set to $h=3$, $k=1$, $l=5$ and
symmetrical exchanges. The distance of $d=2333\,\mathrm{mm}$, the photon energy
$E_\text{ph}=7970\,\mathrm{eV}$, and the detector area of $170\times180\,\mathrm{mm^2}$ are chosen
in accordance with figure~\ref{fig:example}.}
\label{fig:example_theory}
\end{figure}

The filtered images are then compared to the numerical solution of equation~\eqref{eq:pattern}.
A simulation corresponding to the experimental frame in figure~\ref{fig:example} is shown in
figure~\ref{fig:example_theory}. The photon energy, the distance and the detector area were chosen 
from the experimental conditions. The value of the silicon lattice constant $a=543.108\,\mathrm{pm}$ was
taken from \citeasnoun{mohr2012} and  corrected for thermal expansion to $300\,\mathrm{K}$ using 
the data from \citeasnoun{swenson1983}. The Miller indices were set to $[hkl]=[315]$, and this results 
in a pattern with striking similarity to the experimental image. It is impossible to fulfill the Bragg
condition for any other combination of indices in the considered range at the
given photon energy.

\begin{figure}
\includegraphics[width=0.95\columnwidth]{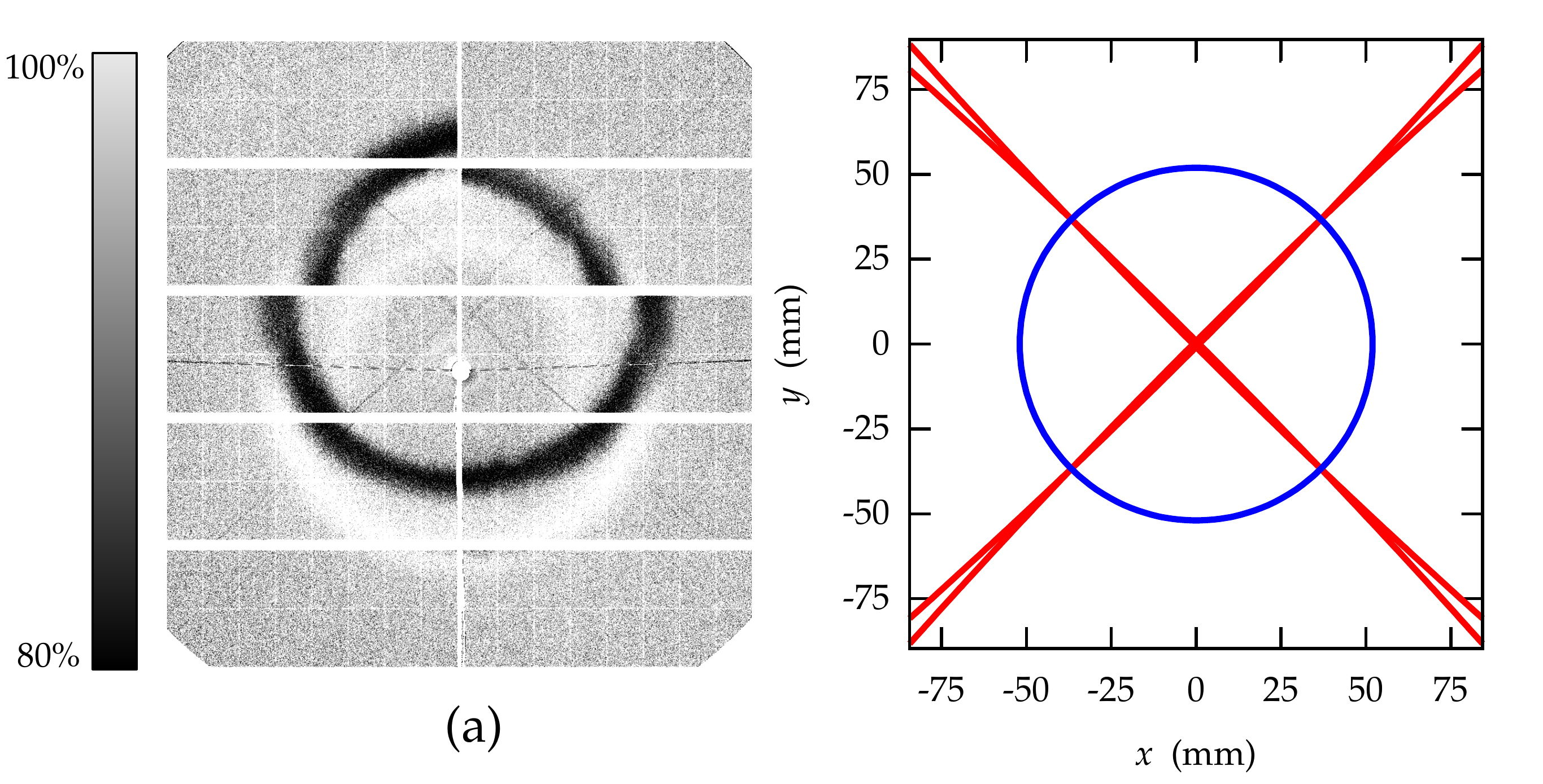}\\
\includegraphics[width=0.95\columnwidth]{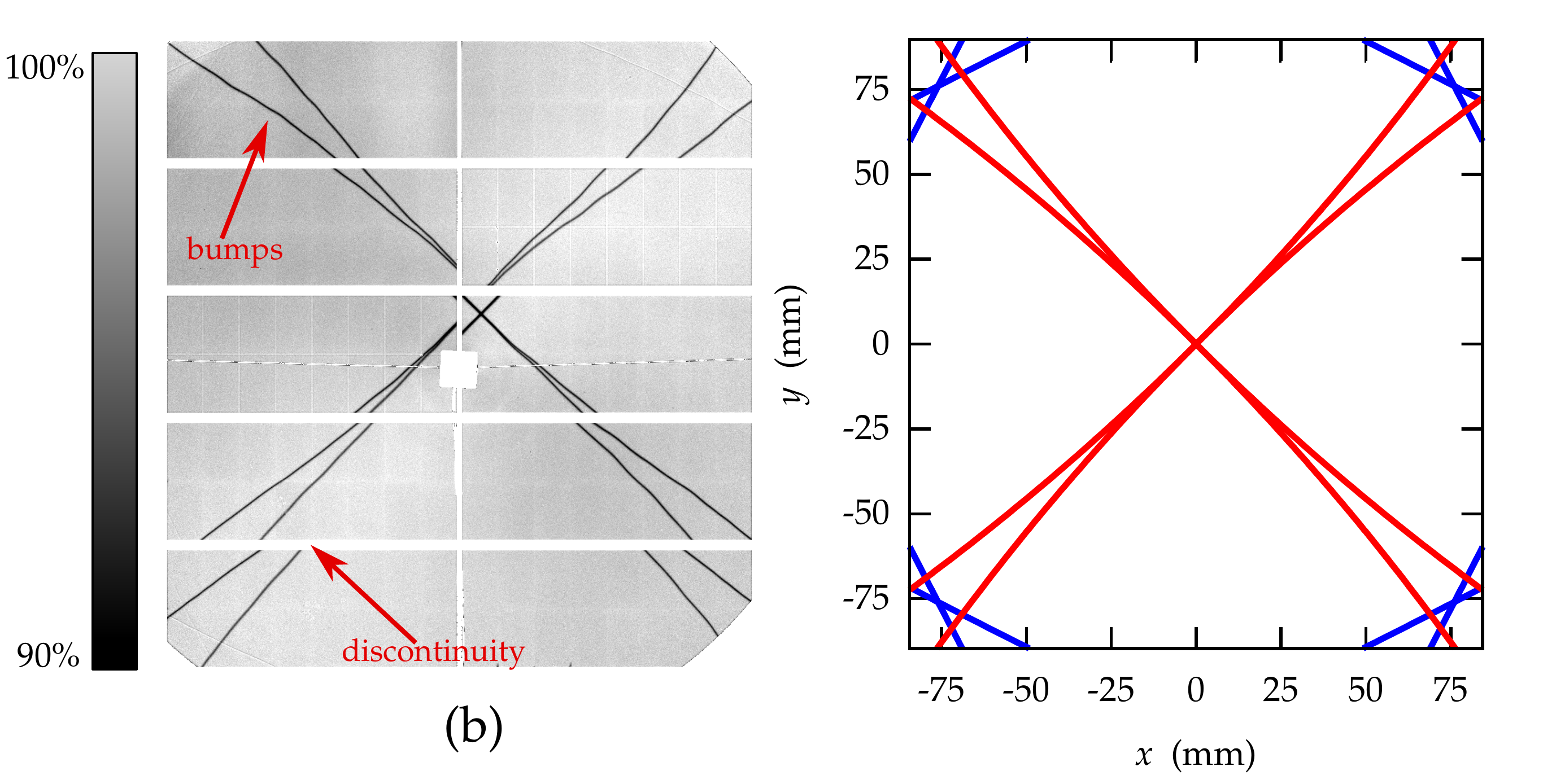}\\
\includegraphics[width=0.95\columnwidth]{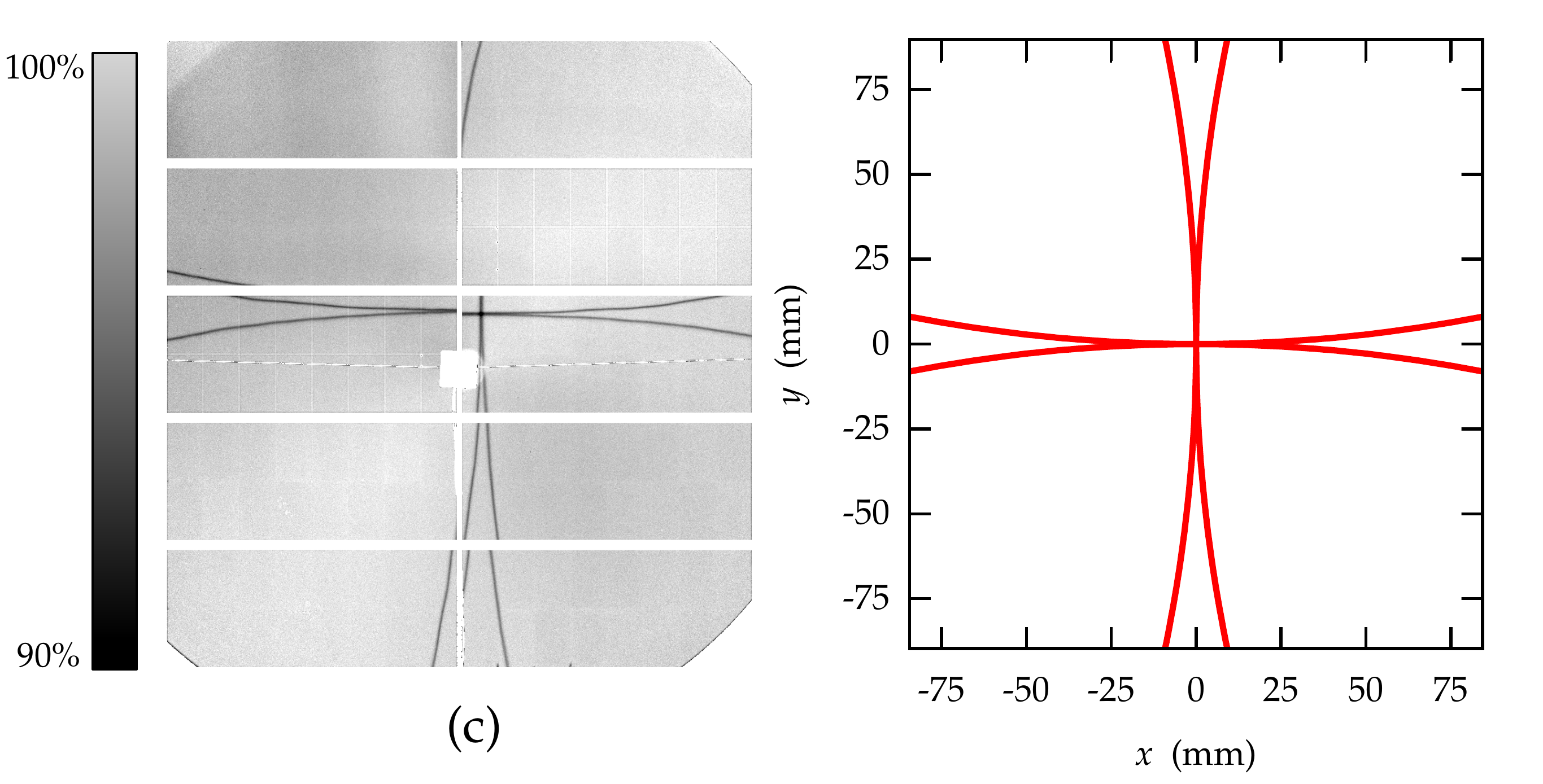}\\
\includegraphics[width=0.95\columnwidth]{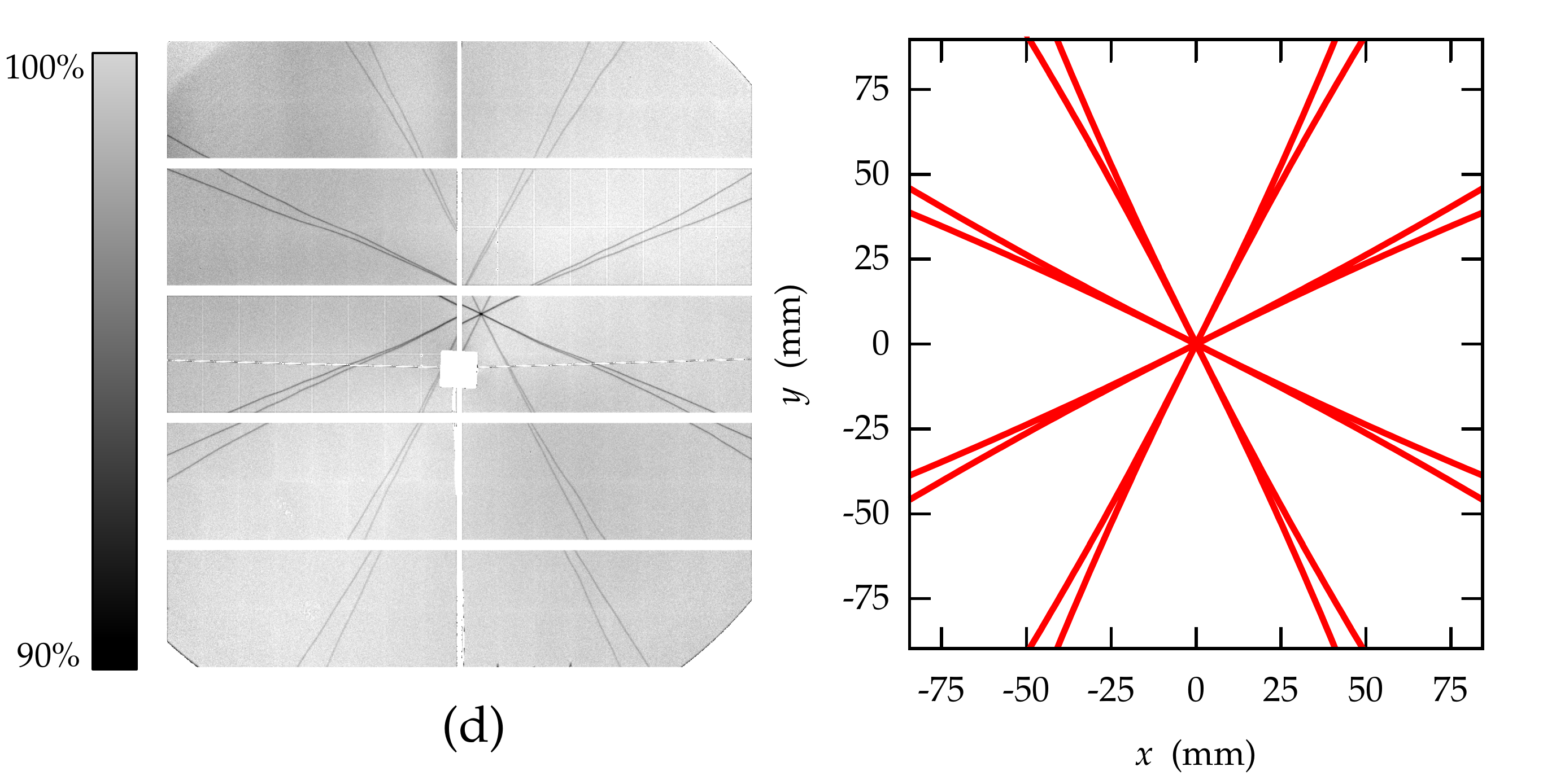}\\
\includegraphics[width=0.95\columnwidth]{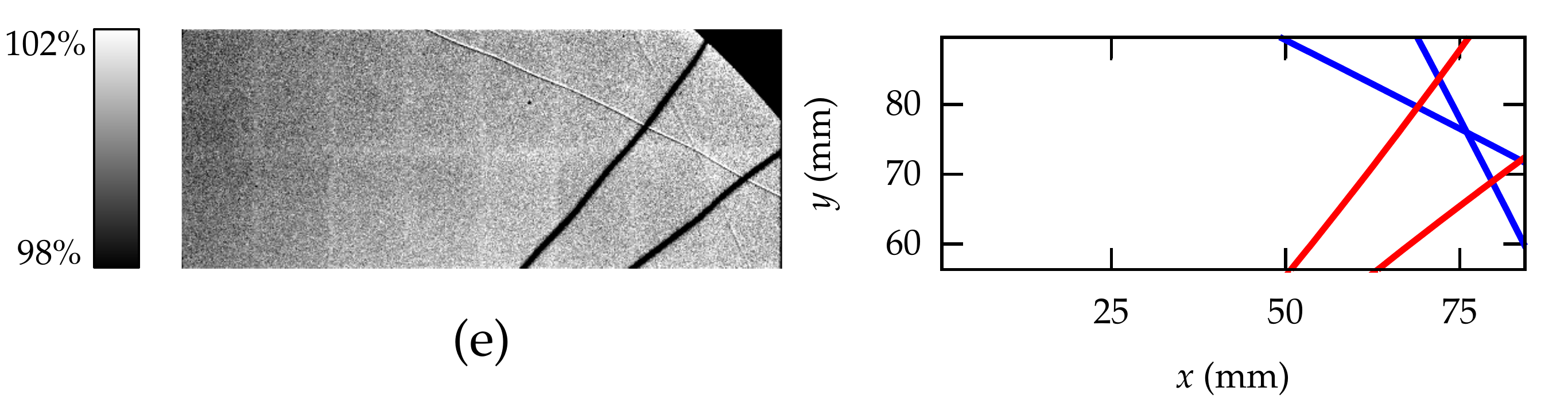}
\caption{Comparison of experimentally obtained patterns with the numerical simulation for
(a) $E_\text{ph}=4567\,\mathrm{eV}, (202)$ in red and $(004)$ in blue,
(b) $E_\text{ph}=7610\,\mathrm{eV}, (206)$ in red and $(313)$ in blue, 
(c) $E_\text{ph}=8316\,\mathrm{eV}, (117)$,
(d) $E_\text{ph}=7990\,\mathrm{eV}, (315)$, respectively.
A contrast-enhanced detail of (b) is shown in (e).
The arrows in (b) indicate the most prominent differences between the experimental images and the
theory. 
A movie containing all recorded images in a range of $E_\text{ph}$ between $7500\,\mathrm{eV}$ and 
$9200\,\mathrm{eV}$ in steps of $5\,\mathrm{eV}$ and around $E_{004}$ can be found in the supplementary
material at \url{http://www.auriocus.de/Video/spinnweb.html}.} 
\label{fig:comparison}
\end{figure}

A comparison of simulated and experimental data at a selection of the characteristic energies $E_{hkl}$ in the
experimentally covered photon energy range is shown in figure~\ref{fig:comparison}.
The possible patterns fall into four categories, each of which is represented by one image in
figure~\ref{fig:comparison}. When both $h=k=0$ (figure
\ref{fig:comparison}a), the expected pattern is a circle with the centre at
the point of normal incidence, which has been discussed already in section~\ref{sec:theory}.
This image was taken at $1.7\,$eV above the characteristic energy, because the circular pattern exists 
only for $E_\text{ph}>E_{hkl}$. For even larger energies, the pattern will be outside
the detector area. This demonstrates the strong sensitivity of the patterns to the photon energy.

When $h \not =0$ and $k=0$, the lines are oriented parallel to the crystalline
coordinate system. For a detector with the $x$-axis parallel to $[110]$, this results in a
diagonal cross, as shown in figure~\ref{fig:comparison}b. A much weaker pattern of the same type is
also present in figure~\ref{fig:comparison}a, which comes from the $(202)$ plane. 

For $h=k\not =0$, the resulting 4-fold symmetry is oriented along the
$[110]$ direction, which displays an upright cross (figure~\ref{fig:comparison}c).
Finally, in the general case $h\not =k\not = 0$, an 8-fold symmetry
is observed (figure~\ref{fig:comparison}d).
All patterns predicted in table~\ref{tbl:ecrit} have been experimentally confirmed up
to a photon energy of $10\,\mathrm{keV}$. 

The measured intensities of the patterns (see table~\ref{tbl:ecrit}) vary greatly from 
pattern to pattern. Most of the patterns lead to a decrease in the intensity of the pixels on the line with
respect to the undistorted signal. An increase in the line intensity with respect to the surroundings
could only be observed for the pattern coming from the $(111)$ plane, and
the patterns caused by reflections at the $(313)$ and $(404)$ planes display a double line
with lower and higher intensity side by side (see figure~\ref{fig:comparison}e for an example).
The former can be explained by reabsorption of the
scattered photons in the same pixel, where the reflection occurs, which improves the absorbance and
thus the quantum efficiency of the detector. The latter is explained by the reabsorption
of the reflected photons in a nearby pixel, thus transferring the measured signal from the dark to
the bright pattern line. 

While the strongest pattern, for instance the $(004)$ circle, leads to a
$20\,\%$ intensity decrease compared to the undistorted signal,
some of the weakest lines approach only a decrease by $\approx0.5\,\%$, as listed in table~\ref{tbl:ecrit}. 
These intensities significantly exceed the noise even for moderate count rates, and may therefore
impede the data evaluation for applications such as protein
crystallography~\cite{uson1999}, if the existence of the pattern is neglected. 
A quantitative theoretical analysis of the line intensities, which requires a space-resolved 
dynamical scattering theory, is beyond the scope of this paper. 

The prediction of the theory is accurate in terms of the energy scale. This can be concluded from
the figures~\ref{fig:comparison}b, c and d, where the photon energy of the exposures is off by less than
$0.5\,\mathrm{eV}$ from the true characteristic energy. The pattern lines meet at one point, as predicted by
equation~\eqref{eq:ecrit}. Therefore the computed characteristic energies agree with the experimentally determined
intersection points of the patterns up to the energy resolution of the monochromator
of around $0.5\,\mathrm{eV}$.

Despite the good agreement in terms of the energy scale, there are some differences 
when the pattern is compared to the simulations in detail, which are marked with arrows
in figure~\ref{fig:comparison}b.
First, large discontinuities disrupt the pattern across the gap between individual
modules. This artefact is especially prominent between the two lowest rows in
figure~\ref{fig:comparison}b. Second, the lines are not ideally smooth, but display bumps,
especially prominent in the top row of figure~\ref{fig:comparison}b. 
The discontinuities arise most probably from an imperfect alignment of the modules with respect to
each other. The angular sensitivity of the pattern can be estimated from the width
of the lines, which is approximately $90$ arc seconds corresponding to $1\,\mathrm{mm}$ in
the given setup. Therefore, even the slightest angular deviation manifests itself as a shift in the pattern.
Another possible reason is the imperfect orientation of the sensor surface with respect to the
crystalline coordinate system, which may vary between individual modules. 
Similarly, the bumps in the lines probably come from the roughness of the detector surface. This roughness
may stem from either the production process of the silicon wafer or from mechanical stress
in the wafer, which is permanently bump-bonded to the supporting circuitry. It should be noted that the deviations
visible here are only observed in the Bragg line pattern and not in the primary scattering image,
which is rather insensitive to the angular misalignment. 

\section{Potential Applications}
\label{sec:Applications}

Due to the sensitivity of the Bragg line patterns to the energy and angle of incidence, a number of
applications can be considered. It is straightforward to check the energy calibration of
the monochromator with the aid of the characteristic energies $E_{hkl}$. When the photon energy is set to one of
the characteristic energies, which are well distributed in a large range (see table~\ref{tbl:ecrit}), all
the lines must meet at one point. If this is not the case, the sign of the deviation can be
concluded from the curvature of the lines. When the inner polygon formed by the lines is concave,
the photon energy is below the characteristic energy, and vice versa, for a convex inner polygon. This
follows from the fact that the radius of the curved lines increases with the energy. In principle,
this method even allows measurement of the energy over a certain range in the neighbourhood of $E_{hkl}$,
by determining the area of the inner polygon. For conclusive results, however, it is necessary
to know the distance from the source, which may limit the accuracy.

Another possible application is the determination of the angular alignment. The point at which the lines
meet, i.e.\ the normal incidence on the detector plane, should coincide with the direct beam for a
perfectly aligned detector. In the experimental data shown in section~\ref{sec:results},
this is clearly not the case. However, the distance 
of the pattern centre from the direct beam position is only $16\,\mathrm{mm}$, which corresponds to an
angular misalignment of $0.4$\textdegree. For the application of recording scattering images, this is
negligible, since the cosine of this angle differs by only one part in $40,000$ from unity. The discontinuity of
the pattern across module borders and the height of the line bumps can be evaluated in a similar way to compute the 
angle between individual modules and the roughness of the detector surface. The result is of a
similar scale, and thus the deviations are too small to be observed in regular scattering images.

\section{Conclusion} 
\label{sec:Conclusion}

A line pattern in X-ray detector images has been discovered which results from Bragg scattering in the
sensor layer of the detector. The pattern overlays all images recorded in typical small-angle scattering
geometry when the sensor layer of the detector is made of crystalline material, which is the case for
all state-of-the-art hybrid pixel detectors. The intensity of this pattern is sufficient 
to disturb the evaluation of scattering and diffraction experiments. First theoretical considerations can explain the
observed patterns with perfect agreement in the investigated photon energy range. The effect can be
exploited to accurately measure the photon energy and the angular alignment of the detector with the
primary beam.


\ack{The experimental contributions from Stefanie Marggraf and Levent Cibik are gratefully
acknowledged. We thank Frank Scholze and Jan Wernecke for interesting discussions. We would also like to thank
Armin Hoell for the research cooperation with the HZB SAXS instrument as well as DECTRIS for
providing data and support with the PILATUS detector.}


\referencelist[spinnweb]
\end{document}